\documentclass[12pt]{article}
\usepackage{graphicx}
\usepackage{xspace}

\newcommand\pubnumber{BABAR-PROC-10/092}
\newcommand\pubdate{\today}

\def\lpnhe{Laboratoire de Physique Nucl\'eaire et de Hautes Energies\\
  IN2P3/CNRS, F-75252 Paris, FRANCE}
\def\support{\footnote{e-mail: giovanni.marchiori@lpnhe.in2p3.fr}}

\def\Title#1{\begin{center} {\Large #1 } \end{center}}
\def\Author#1{\begin{center}{ \sc #1} \end{center}}
\def\Address#1{\begin{center}{ \it #1} \end{center}}

\newcommand\pubblock{\rightline{\begin{tabular}{l} \pubnumber\\
         \pubdate  \end{tabular}}}
\newenvironment{Abstract}{\begin{quotation}  }{\end{quotation}}
\newenvironment{Presented}{\begin{quotation} \begin{center} 
             PROCEEDINGS OF\end{center}\bigskip 
      \begin{center}\begin{large}}{\end{large}\end{center} \end{quotation}}

\def\beq{\begin{equation}}
\def\eeq#1{\label{#1}\end{equation}}
\def\eeqn{\end{equation}}

\def\beqa{\begin{eqnarray}}
\def\eeqa#1{\label{#1}\end{eqnarray}}
\def\eeqan{\end{eqnarray}}

\let\bar=\overbar

\def\Dslash{\not{\hbox{\kern-4pt $D$}}}
\def\dslash{\not{\hbox{\kern-2pt $\del$}}}

\def\msb{{\bar{\ssstyle M \kern -1pt S}}}

\def\Bbar    {\kern 0.18em\overline{\kern -0.18em B}{}\xspace} 
\def\BB      {\ensuremath{B\Bbar}\xspace}  
\def\KS    {\ensuremath{K^0_{\scriptscriptstyle S}}\xspace}
\def\CP {\ensuremath{C\!P}\xspace}
\def\de {\mbox{$\Delta E$}\xspace} 
\def\mes{\mbox{$m_{\rm ES}$}\xspace}
\def\degrees{\ensuremath{^{\circ}}\xspace}
\newcommand{\fis}{\ensuremath{\mbox{$\mathcal{F}$}}\xspace} 
\newcommand{\gevcc}{\ensuremath{{\mathrm{\,Ge\kern -0.1em V\!/}c^2}}\xspace} 

\def\AcppVal{0.25}
\def\AcppErrStat{0.06}
\def\AcppErrSyst{0.02}
\def\AcpmVal{-0.09}
\def\AcpmErrStat{0.07}
\def\AcpmErrSyst{0.02}
\def\RcppVal{1.18}
\def\RcppErrStat{0.09}
\def\RcppErrSyst{0.05}
\def\RcpmVal{1.07}
\def\RcpmErrStat{0.08}
\def\RcpmErrSyst{0.04}

\def\dgLoAi{11.3}	
\def\dgHiAi{22.7}
\def\dgLoAj{80.9}
\def\dgHiAj{99.1}
\def\dgLoAk{157.3}
\def\dgHiAk{168.7}

\def\dgLoBi{7.0}	
\def\dgHiBi{173.0}

\def\rLoA{0.24}		
\def\rHiA{0.45}

\def\rLoB{0.06}		
\def\rHiB{0.51}

\def\xpValNoKsPhi{-0.057}	
\def\xpErrStatNoKsPhi{0.039}
\def\xpErrSystNoKsPhi{0.015}

\def\xmValNoKsPhi{0.132}	
\def\xmErrStatNoKsPhi{0.042}
\def\xmErrSystNoKsPhi{0.018}

\def\Lbabar{\mbox{{\fontsize{18}{11}\sl B}\hspace{-0.05em}{\fontsize{14}{11}\sl A}\hspace{-0.07em}{\fontsize{18}{11}\sl B}\hspace{-0.05em}{\fontsize{14}{11}\sl A\hspace{-0.02em}R}}}
\def\lbabar{\mbox{{\fontsize{12}{11}\sl B}\hspace{-0.05em}{\fontsize{10}{11}\sl A}\hspace{-0.03em}{\fontsize{12}{11}\sl B}\hspace{-0.35em} {\fontsize{10}{11}\sl A\hspace{-0.02em}R}}}
\def\babar{\mbox{\sl B\hspace{-0.37em} {\small\sl A}\hspace{-0.3em}
    \sl B\hspace{-0.37em} {\small\sl A\hspace{-0.02em}R}}}
\begin{document}
\begin{titlepage}
\pubblock

\vfill
\Title{Time-integrated measurements of the CKM angle $\gamma/\phi_3$ in \Lbabar}
\vfill
\Author{Giovanni Marchiori\support\\on behalf of the \babar\ Collaboration}
\Address{\lpnhe}
\vfill
\begin{Abstract}
The most recent determinations of the CKM angle $\gamma/\phi_3$ by the \lbabar\
Collaboration, using time-integrated observables measured in charged 
$B\to D^{(*)}K^{(*)}$ decays, are presented.
The measurements have been performed on the full sample of 468 million 
\BB~pairs collected by the \lbabar\ detector at the SLAC 
PEP-II asymmetric-energy $B$ factory in the years 1999-2007.
\end{Abstract}
\vfill
\begin{Presented}
CKM2010, the 6th International Workshop on the CKM Unitarity Triangle\\
University of Warwick, UK\\
6-10 September 2010
\end{Presented}
\vfill
\end{titlepage}
\def\thefootnote{\fnsymbol{footnote}}
\setcounter{footnote}{0}

\section{Introduction}
A theoretically clean measurement of the angle 
$\gamma\equiv\arg \left[-\frac{V_{ud}V_{ub}^*}{V_{cd}V_{cb}^*}\right]$ 
(also denoted as $\phi_3$ in the literature)
can be obtained using \CP-violating $B\to D^{(*)}K^{(*)}$ decays. 
The interference between the $b\to c\bar u s$ and
$b\to u\bar c s$ tree amplitudes results in observables that depend
on the relative weak phase $\gamma$, the magnitude ratio $r_B\equiv\left|
\frac{A(b\to u)}{A(b\to c)}\right|$, and the relative strong phase
$\delta_B$ between the two amplitudes.
The hadronic parameters, $r_B$ and $\delta_B$, depend on the $B$ decay
under investigation; they can not be precisely calculated from theory,
but can be extracted directly from data by simultaneously
reconstructing several different $D$ final states.

In this contribution we present the most recent $\gamma$
determinations obtained by \babar, based on the full sample ($\approx
468\times 10^6$ $B^\pm$ decays) of charged $B$ mesons produced in
$e^+e^-\to\Upsilon(4S)\to B^+B^-$ and accumulated in the years
1999-2007. The following decays have been reconstructed: 
(i) $B^\pm\to D^{(*)}K^\pm$ and $B^\pm\to D K^{*\pm}
(K^{*\pm}\to \KS\pi^\pm)$, with $D\to \KS h^+h^-$,
$h=\pi,K$;
(ii) $B^\pm\to DK^\pm$, with $D$ decaying to \CP-eigenstates $f_{\CP}$;
(iii) $B^\pm\to D^{(*)}K^\pm$, with $D$ decaying to $K^\pm\pi^\mp$.
The results are statistically limited, as the effects that
are being searched for are tiny, since: (i)
the branching fractions of the $B$ meson decays considered here
are on the order of $5\times 10^{-4}$ or lower; (ii) the
branching fractions for $D^{(*)}$ decays, including secondary decays,
range between $O(10^{-2})$ and $O(10^{-4})$; (iii)
the interference between the $b\to c$ and $b\to u$
mediated $B$ decay amplitudes is low, as the ratios
$r_B$ are around 0.1 due to CKM factors and the additional
color-suppression of $A(b\to u)$.

The $B$ decay final states are completely reconstructed,
with efficiencies between 40\% (for low-multiplicity,
low-background decay modes) and 5\% (for high-multiplicity decays).
The selection is optimized to maximise the statistical sensitivity
$S/\sqrt{S+B}$, where the number of expected signal ($S$) and background
($B$) events is estimated from simulated samples and data control samples.
Signal $B$ decays are distinguished from \BB and continuum $q\bar{q}$
background by means of maximum likelihood fits to two variables exploiting 
the kinematic constraint from the known beam energies: the energy-substituted
invariant mass $\mes{\equiv}\sqrt{E^{*2}_{\rm beam}-p_B^{*2}}$ and the 
energy difference 
$\Delta E{\equiv}E_B^*{-}E^{*}_{\rm beam}$. Additional continuum background
discrimination is achieved by including in the likelihood a variable
built, using multivariate analysis tools, from the combination (either a linear
Fisher discriminant, \fis, or a non-linear neural-network, $NN$) of several
event-shape quantities. These variables distinguish spherical \BB events from
more jet-like $q\bar{q}$ events and exploit the different angular correlations
in the two event categories. $B\to D^{(*)}\pi$ decays, which are 12 times more 
abundant than $B\to D^{(*)}K$ and are expected to show negligible \CP-violating 
effects ($r_B\approx 0.01$ in such decays), are discriminated by means of the 
excellent pion and kaon identification provided by $dE/dx$ measured in the 
charged particle tracking devices and by the radiation detected in the 
Cherenkov detector, and are used as control samples.

\section{Dalitz-plot method: $B^\pm\to D^{(*)}K^{(*)\pm}$, $D\to \KS h^+h^-$}
\begin{figure}[!htb]
\centering
\includegraphics[width=0.32\columnwidth]{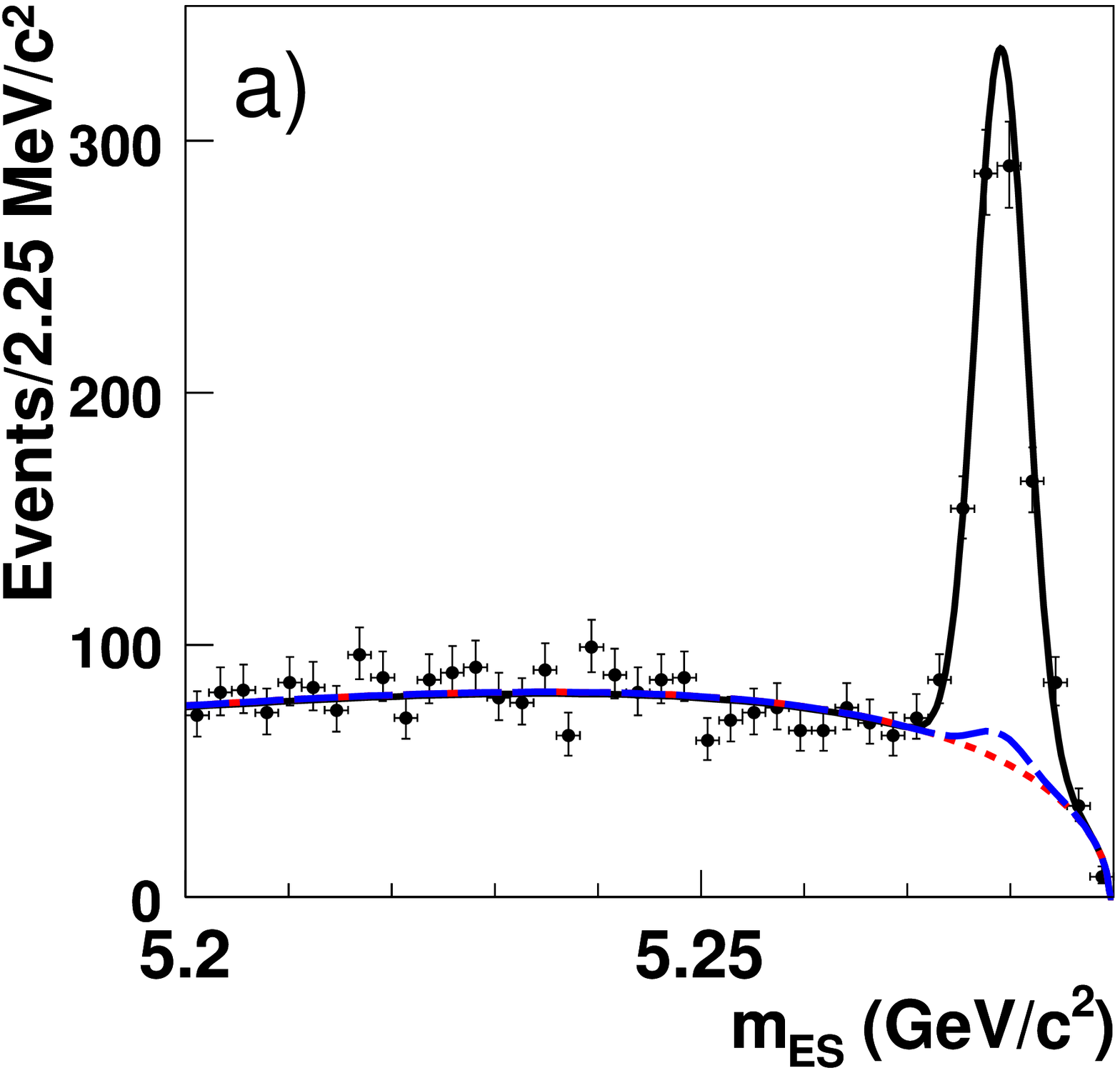}
\includegraphics[width=0.32\columnwidth]{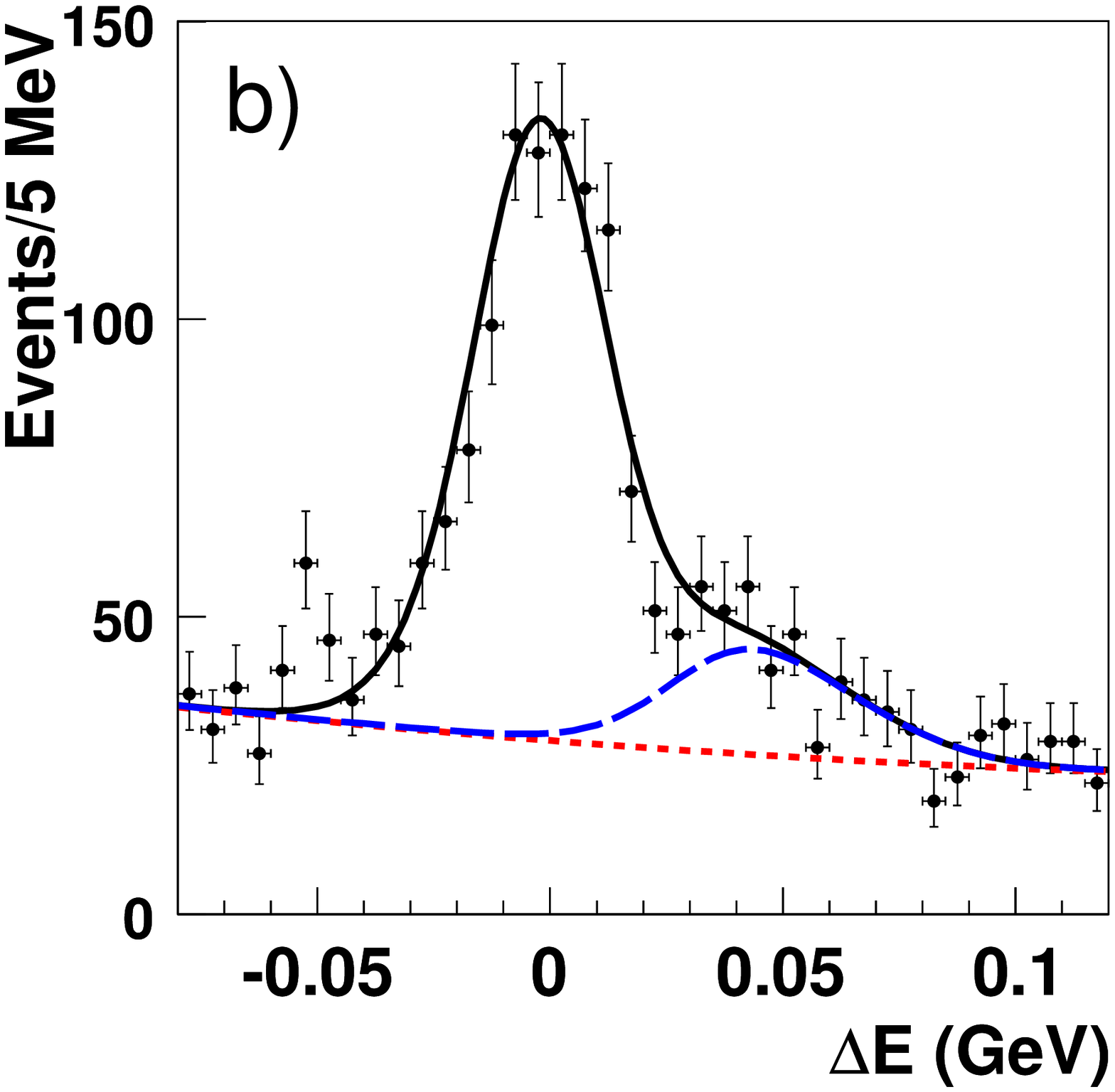}
\includegraphics[width=0.32\columnwidth]{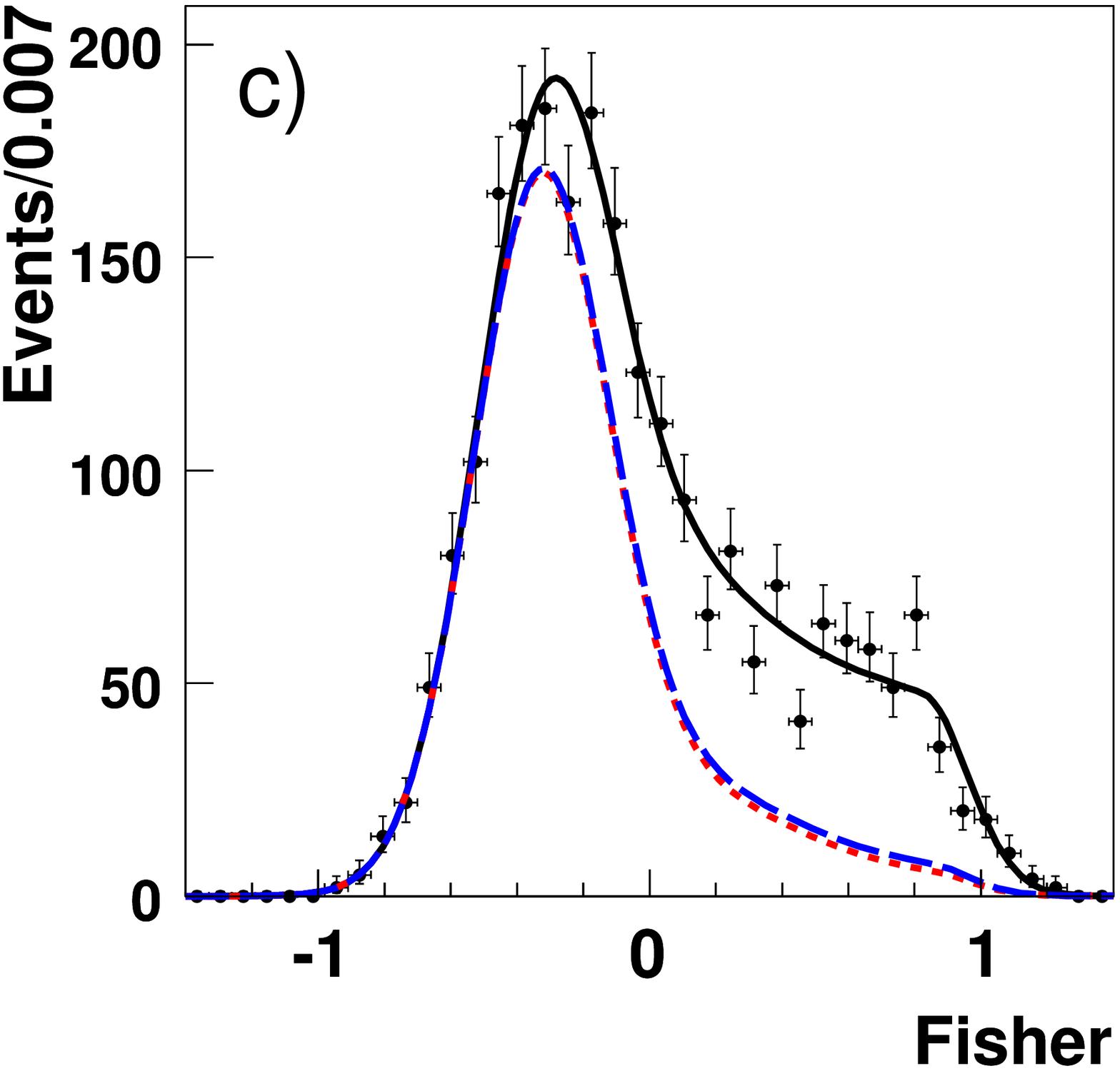}
\caption{The \mes (a), \de (b), and \fis (c)
distributions for $B^\pm{\to}D K^\pm$, $D{\to}\KS \pi^+\pi^-$. 
for events in the signal region ($\mes>5.272$~\gevcc, 
$|\de|<30$ MeV, and $\fis>-0.1$), after all the selection criteria,
except the one on the plotted variable, are applied.
The curves represent the fit projections:  
signal plus background (solid black lines), $q\bar{q}+\BB$
background (dotted red lines), $q\bar{q}+\BB+B\to D\pi$
background (dashed blue lines).}
\label{fig:GGSZ_dk_kspipi_yield_fit}
\end{figure}

We reconstruct $B^\pm{\to}DK^\pm$, $D^{*}K^\pm$ ($D^{^*}{\to}D\gamma$ and
$D\pi^0$), and $DK^{*\pm}$ ($K^{*\pm}{\to}K^0_S\pi^\pm$) decays,
followed by neutral $D$ meson decays to the 3-body self-conjugate
final states $K^0_Sh^+h^-$ ($h=\pi,K$)~\cite{bib:babar_DALITZ}.
From an extended maximum likelihood fit to \mes, \de and \fis 
(Fig.~\ref{fig:GGSZ_dk_kspipi_yield_fit})
we determine the signal and background yields in each channel:
we find 268 $B$ candidates with $D\to K^0_SK^+K^-$ and
1507 $B$ candidates with $D\to K^0_S\pi^+\pi^-$.

\begin{figure}[!htb]
\centering
\includegraphics[width=0.32\columnwidth]{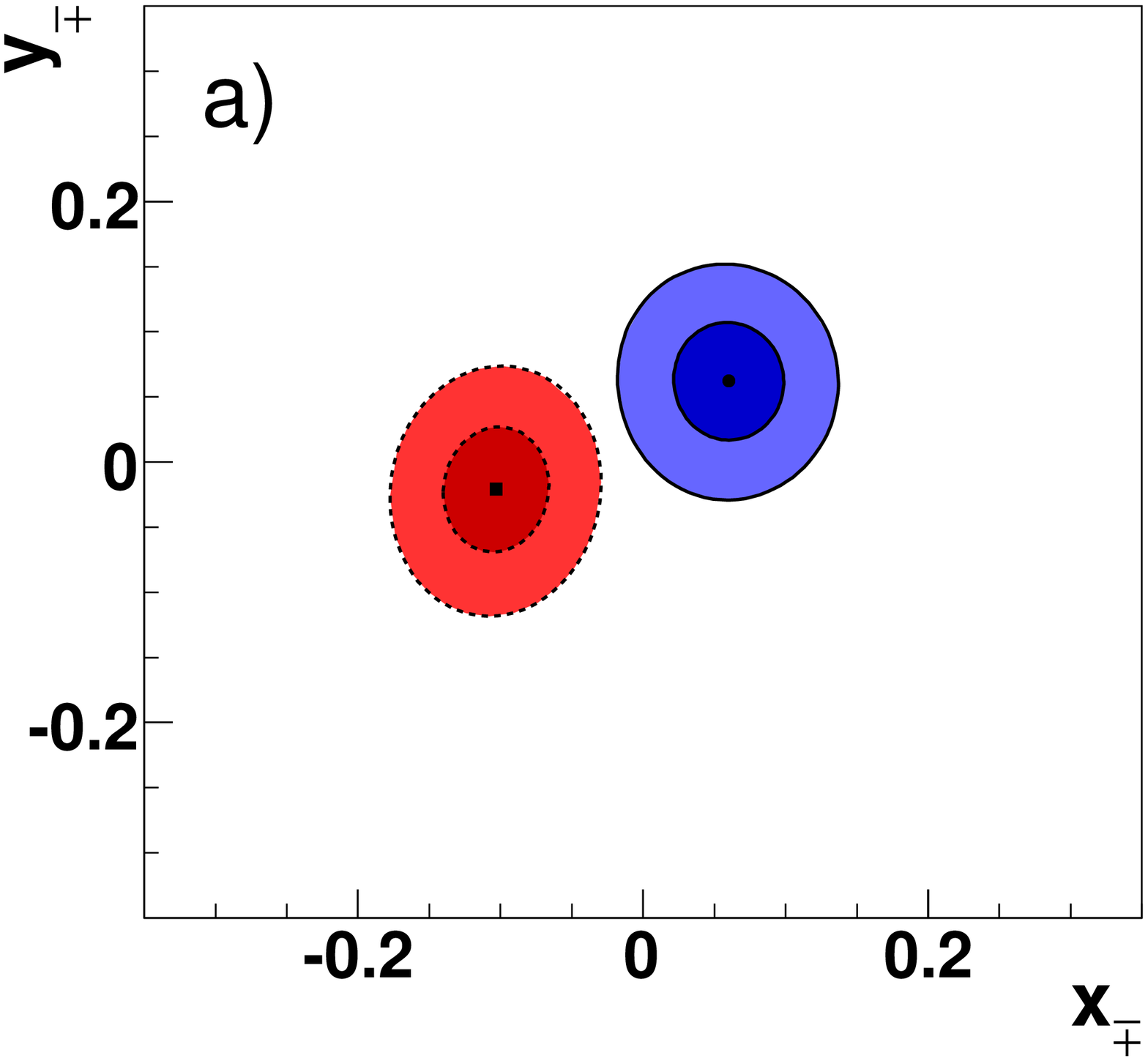}
\includegraphics[width=0.32\columnwidth]{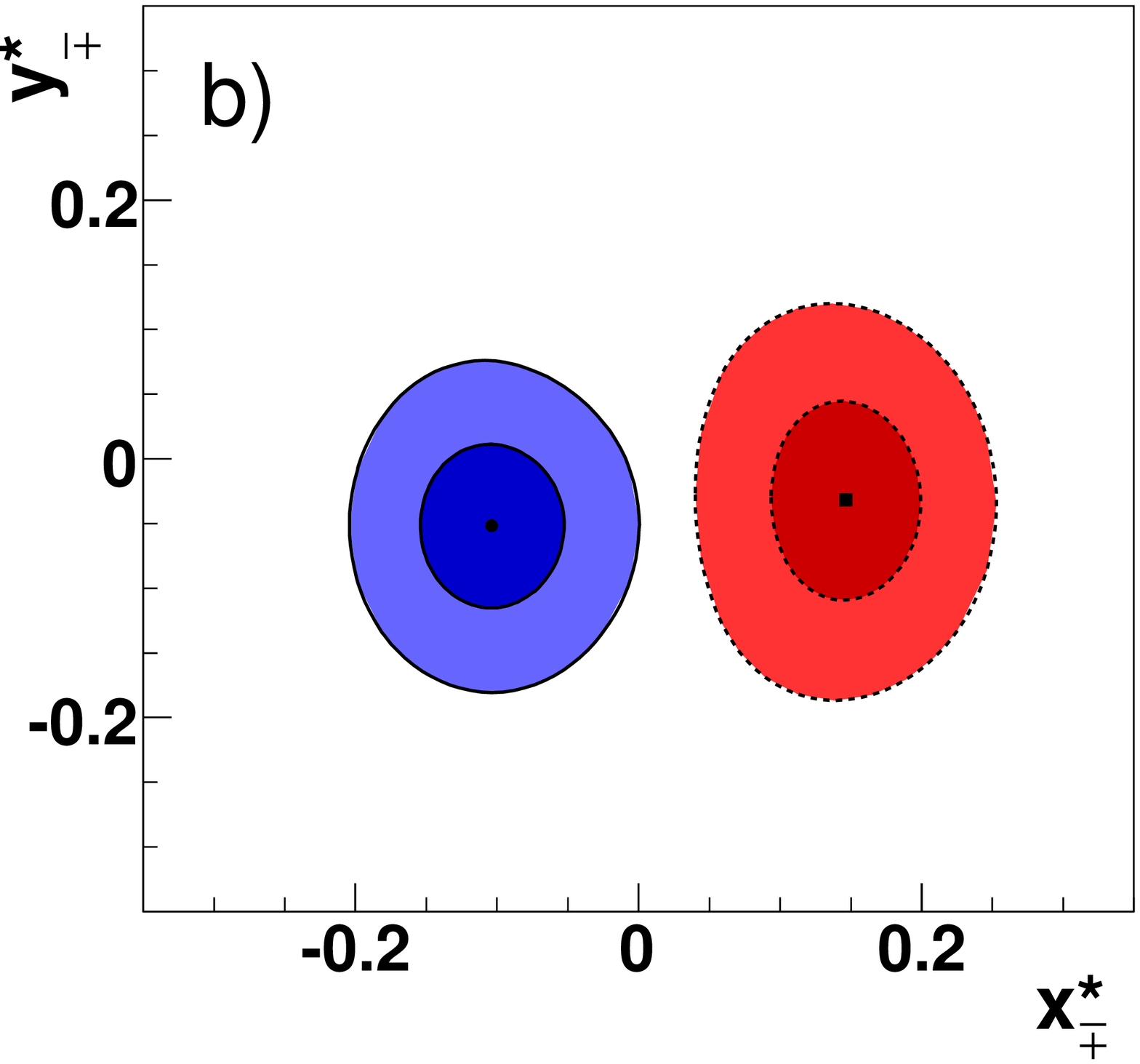}
\includegraphics[width=0.32\columnwidth]{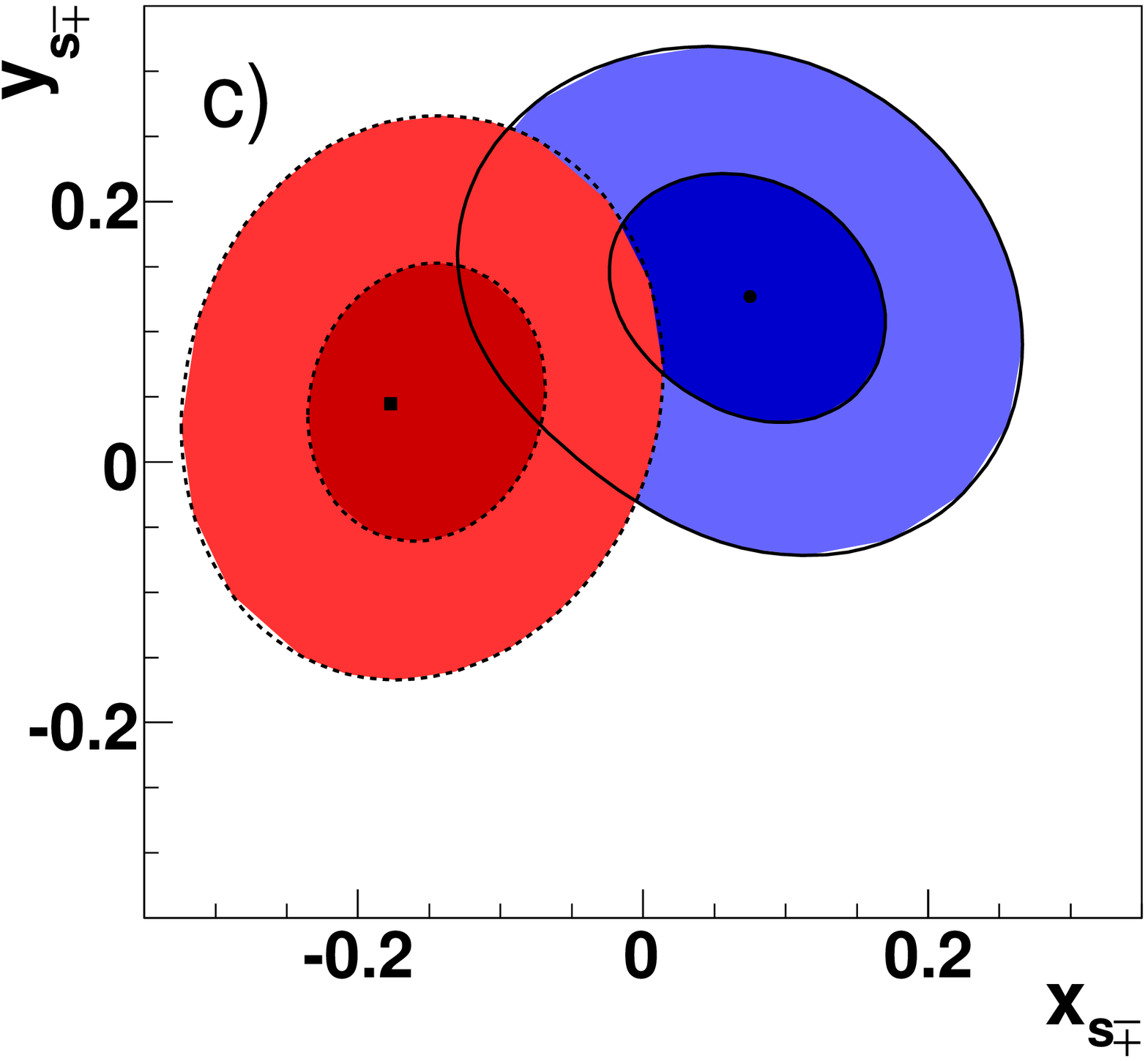}
\caption{$1\sigma$ and $2\sigma$ contours in the ${x_\pm, y_\pm}$
planes for (a) $B\to DK$, (b) $B\to D^*K$ and (c) $B\to DK^*$,
for $B^-$ (solid lines) and $B^+$ (dotted lines) decays.}
\label{fig:GGSZ_cartcoord}
\end{figure}

Following the technique proposed in \cite{bib:GGSZ},
from a fit to the Dalitz-plot distribution of
the $D$ daughters
% (an example is shown in Fig.~\ref{fig:GGSZ_dk_kspipi_dalitz}), 
we determine 2D confidence regions for the
variables $x_\pm \equiv{r_B\cos(\delta_B\pm\gamma)}$ and
$y_\pm\equiv{r_B\sin(\delta_B\pm\gamma)}$ (Fig.~\ref{fig:GGSZ_cartcoord}).
In the fit we model the $D^0$ and $\overline{D}^0$
decay amplitudes to $K^0_Sh^+h^-$ as the coherent sum of
a non-resonant part and several intermediate two-body decays that
proceed through known $K^0_Sh$ or $h^+h^-$ resonances. The
model is determined from large ($\approx 6.2\times 10^5$) and very
pure ($\approx 99\%$) control samples of $D$ mesons produced in
$D^*{\to}D\pi$ decays~\cite{bib:Dmixing}.
%\begin{figure}[!htb]
%\centering
%\includegraphics[width=0.4\columnwidth]{GGSZ_dk_kspipi_dalitz_bminus}
%\includegraphics[width=0.4\columnwidth]{GGSZ_dk_kspipi_dalitz_bplus}
%\caption{Dalitz plot distribution of the $D$ daughters in (a) $B^-\to DK^-$ 
%and $B+\to DK^+$, with $D\to\KS\pi^+\pi^-$. 
%Only events in the signal region are 
%shown. The contours (solid red lines) represent the kinematical limits of 
%the $D$ decay.}
%\label{fig:GGSZ_dk_kspipi_dalitz}
%\end{figure}
The results for $x$ and $y$ are summarized in Table~\ref{tab:xy}.
%They are consistent with and have similar precision to those obtained
%by the Belle experiment.~\cite{bib:belle_DALITZ}
\begin{table}[!h]
\begin{center}
\scriptsize
\begin{tabular}{l|ccc}
\textbf{Parameter} & $B^\pm\to DK^\pm$ & $B^\pm\to D^{*}K^\pm$ & $B^\pm\to DK^{*\pm}$ \\
\hline
$x_{+}$ & $-0.103\pm 0.037\pm 0.006\pm 0.007$           & $\phantom{-}0.147\pm 0.053\pm 0.017\pm 0.003$ & $-0.151\pm 0.083\pm 0.029\pm 0.006$ \\
$y_{+}$ & $-0.021\pm 0.048\pm 0.004\pm 0.009$           & $-0.032\pm 0.077\pm 0.008\pm 0.006$ & $\phantom{-}0.045\pm 0.106\pm 0.036\pm 0.008$ \\
$x_{-}$ & $\phantom{-}0.060\pm 0.039\pm 0.007\pm 0.006$ & $-0.104\pm 0.051\pm 0.019\pm 0.002$ & $\phantom{-}0.075\pm 0.096\pm 0.029\pm 0.007$ \\
$y_{-}$ & $\phantom{-}0.062\pm 0.045\pm 0.004\pm 0.006$ & $-0.052\pm 0.063\pm 0.009\pm 0.007$ & $\phantom{-}0.127\pm 0.095\pm 0.027\pm 0.006$ \\
\hline
\end{tabular}
\caption{Values of $x_\pm$ and $y_\pm$ measured with the Dalitz-plot 
  analysis of $B^\pm{\to}D^{(*)}K^{(*)\pm}$}
\label{tab:xy}
\end{center}
\end{table}

\begin{figure}[!htb]
\centering
\includegraphics[width=0.32\columnwidth]{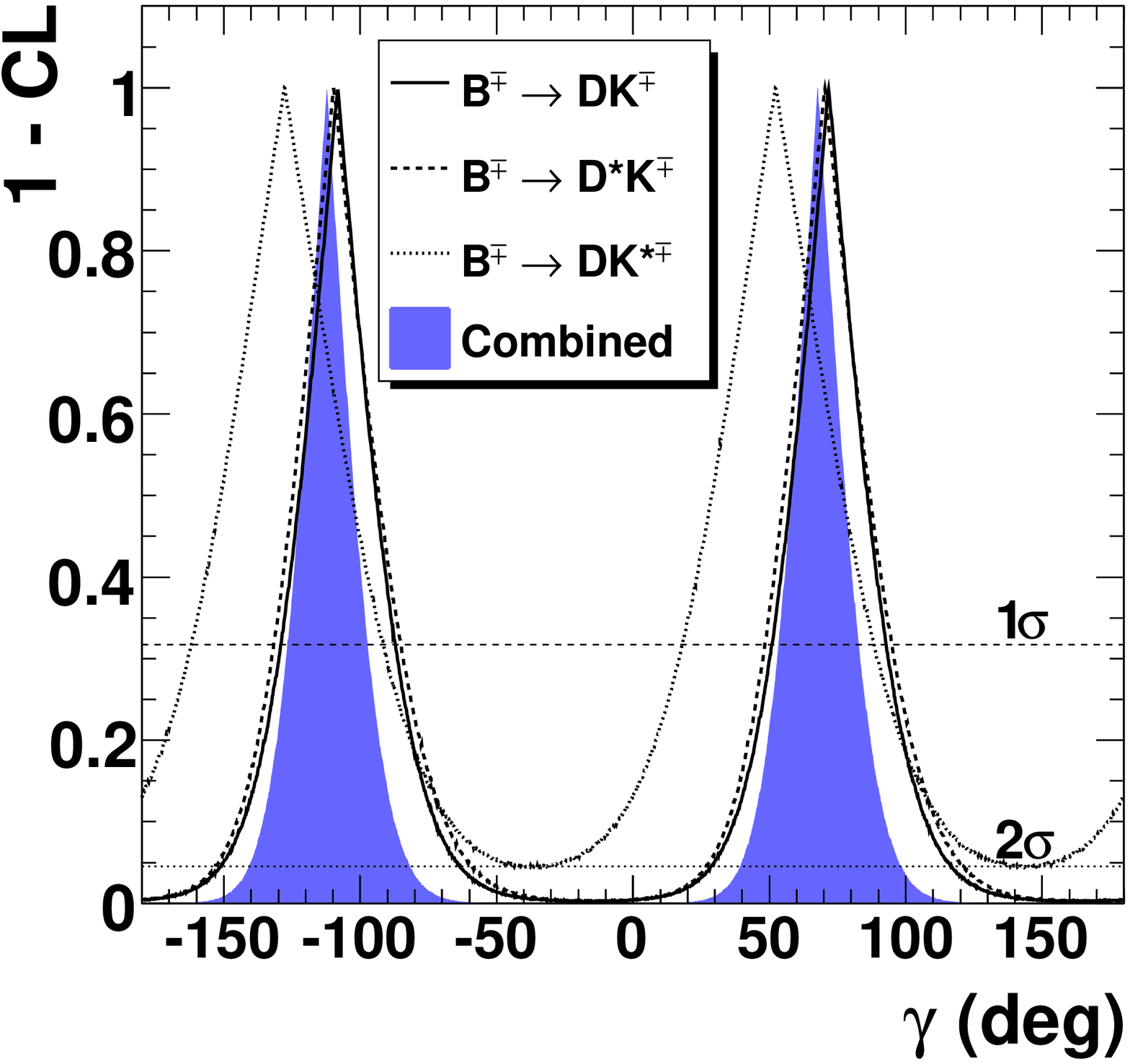}
\includegraphics[width=0.32\columnwidth]{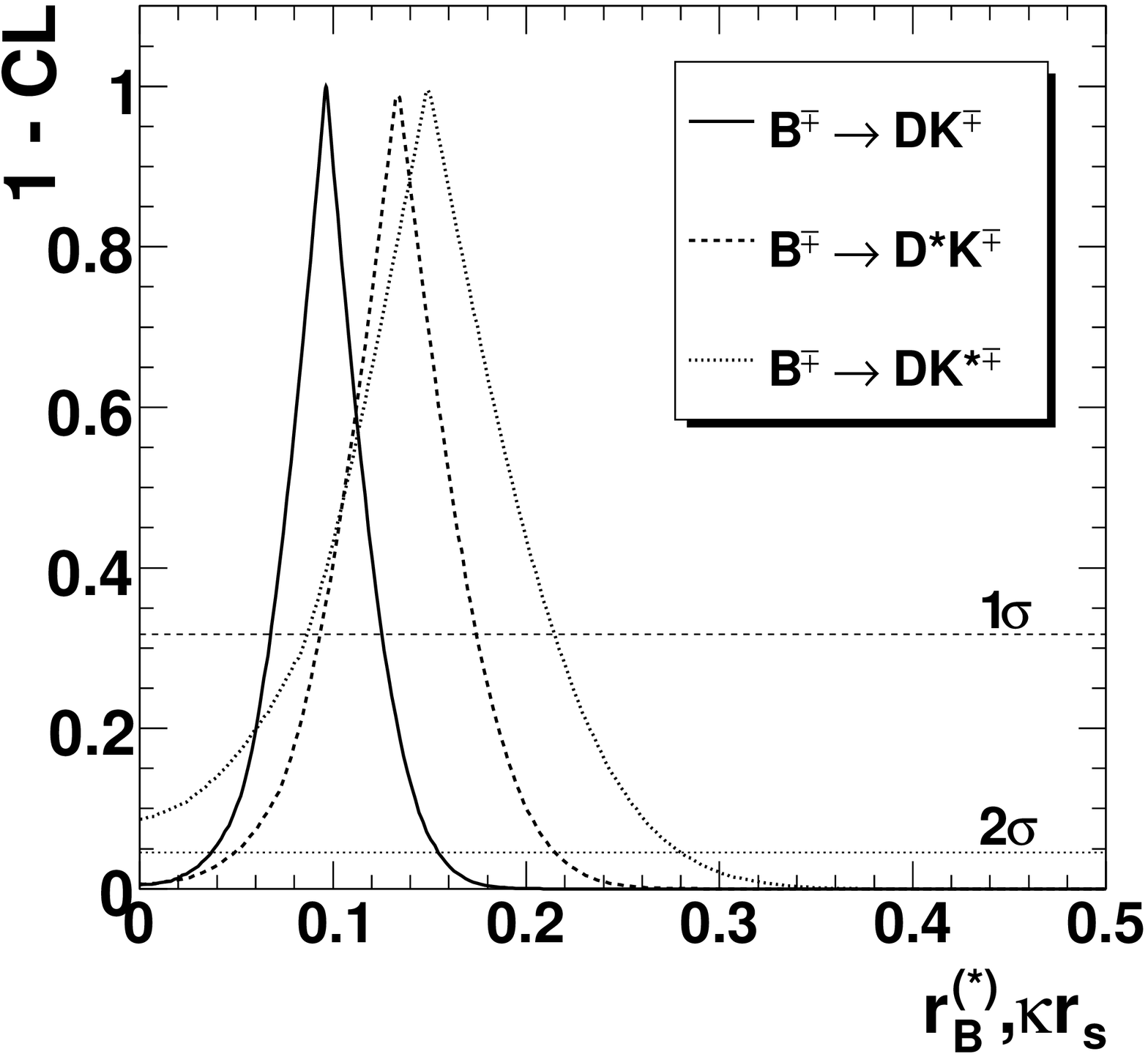}
\includegraphics[width=0.32\columnwidth]{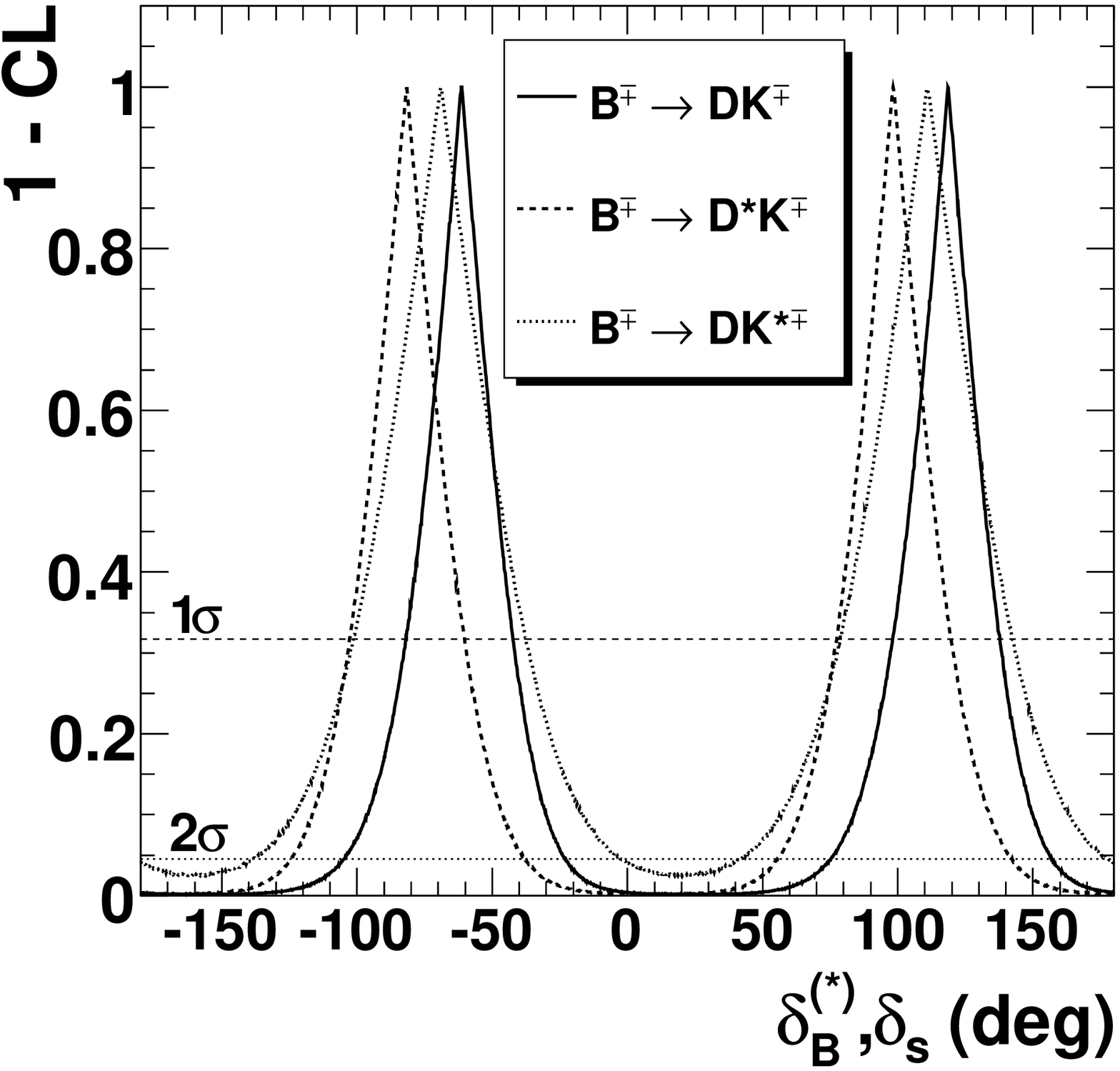}
\caption{1-confidence level (CL) as a function of $\gamma$ (left), $r_B$ (center) and $\delta_B$ (right) from the $B\to D^{(*)}K^{(*)}$ Dalitz-plot analysis.}
\label{fig:GGSZ_r_and_gamma}
\end{figure}

From the $(x_\pm,y_\pm)$ confidence regions we determine, using
a frequentist procedure, $1\sigma$ confidence intervals
for $\gamma$, $r_B$ and $\delta_B$ (Fig.~\ref{fig:GGSZ_r_and_gamma}).
We obtain $\gamma\ {\rm mod}\ 180^\circ= (68\pm 14\pm 4\pm 3)^\circ$,
where the three uncertainties are respectively the statistical, the experimental systematic and the Dalitz-model systematic ones.
We find values of $r_B$ around 0.1, confirming that
interference is low in these channels:
$r^{DK^\pm}_B = 0.096{\pm}0.029$; $r^{D^{*}K^\pm}_B = 0.133^{+0.042}_{-0.039}$;
$kr^{DK^{*\pm}}_B=0.149^{+0.066}_{-0.062}$ ($k{=}0.9{\pm}0.1$ takes into
account the $K^*$ finite width). 
%The small values of $r_B$ favored by our data are responsible - since 
%$\sigma_\gamma{\approx}\sigma_{x,y}/r_B$ - for the larger uncertainty on
%$\gamma$ when compared to the analogous Belle measurement, which favours
%higher values of $r_B$.
We also measure the strong phases (modulo $180^\circ$):
$\delta^{DK^\pm}_B{=}(119^{+19}_{-20})^\circ$;
$\delta^{D^{*}K^\pm}_B{=}(-82\pm 21)^\circ$;
$\delta^{DK^{*\pm}}_B{=}(111\pm 32)^\circ$.
A $3.5\sigma$ evidence of direct \CP\ violation is found
from the distance between $(x_+,y_+)$ and $(x_-,y_-)$ (0
in absence of CPV) in the three $B$ decay channels.

\section{GLW method: $B^\pm\to DK^{(*)\pm}$, $D\to f_{(\rm CP)}$}
We reconstruct $B^\pm\to DK^\pm$ decays, with $D$ mesons
decaying to non-\CP ($D^0\to K^-\pi^+$), \CP-even ($K^+K^-$, $\pi^+\pi^-$) and 
\CP-odd ($\KS\pi^0$, $\KS\phi$, $\KS\omega$) eigenstates~\cite{bib:babar_GLW_d0k}.
The partial decay rate charge asymmetries $A_{\CP\pm}$ for
\CP-even and \CP-odd $D$ final states and the ratios $R_{\CP\pm}$ 
of the charged-averaged $B$ meson partial decay rates in \CP and non-\CP decays
provide a set of four observables from which the three unknowns $\gamma$, $r_B$ and 
$\delta_B$ can be extracted (with an 8-fold discrete ambiguity for the phases)~\cite{bib:GLW}.

\begin{figure}[!htb]
\centering
\includegraphics[width=0.48\columnwidth]{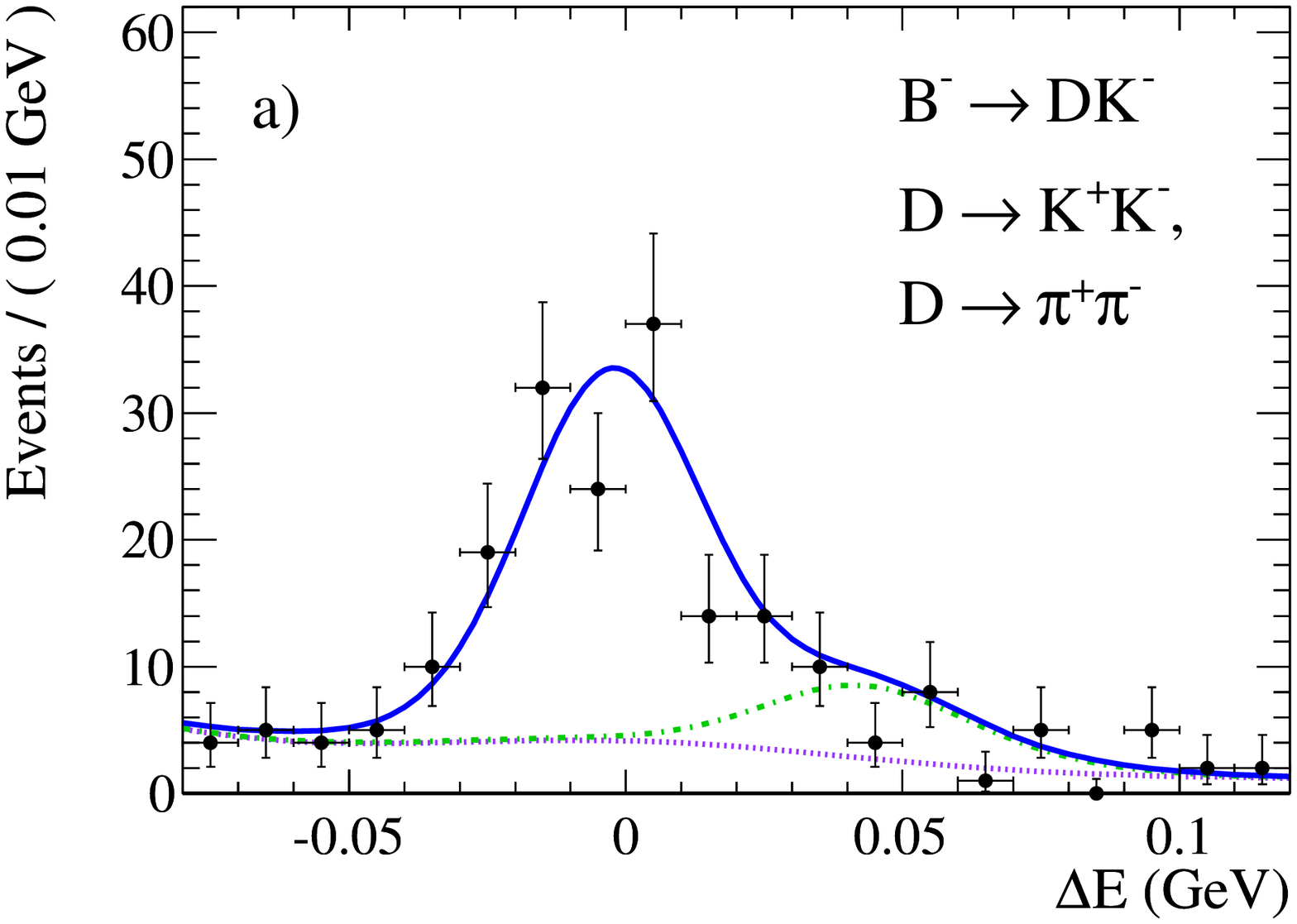}
\includegraphics[width=0.48\columnwidth]{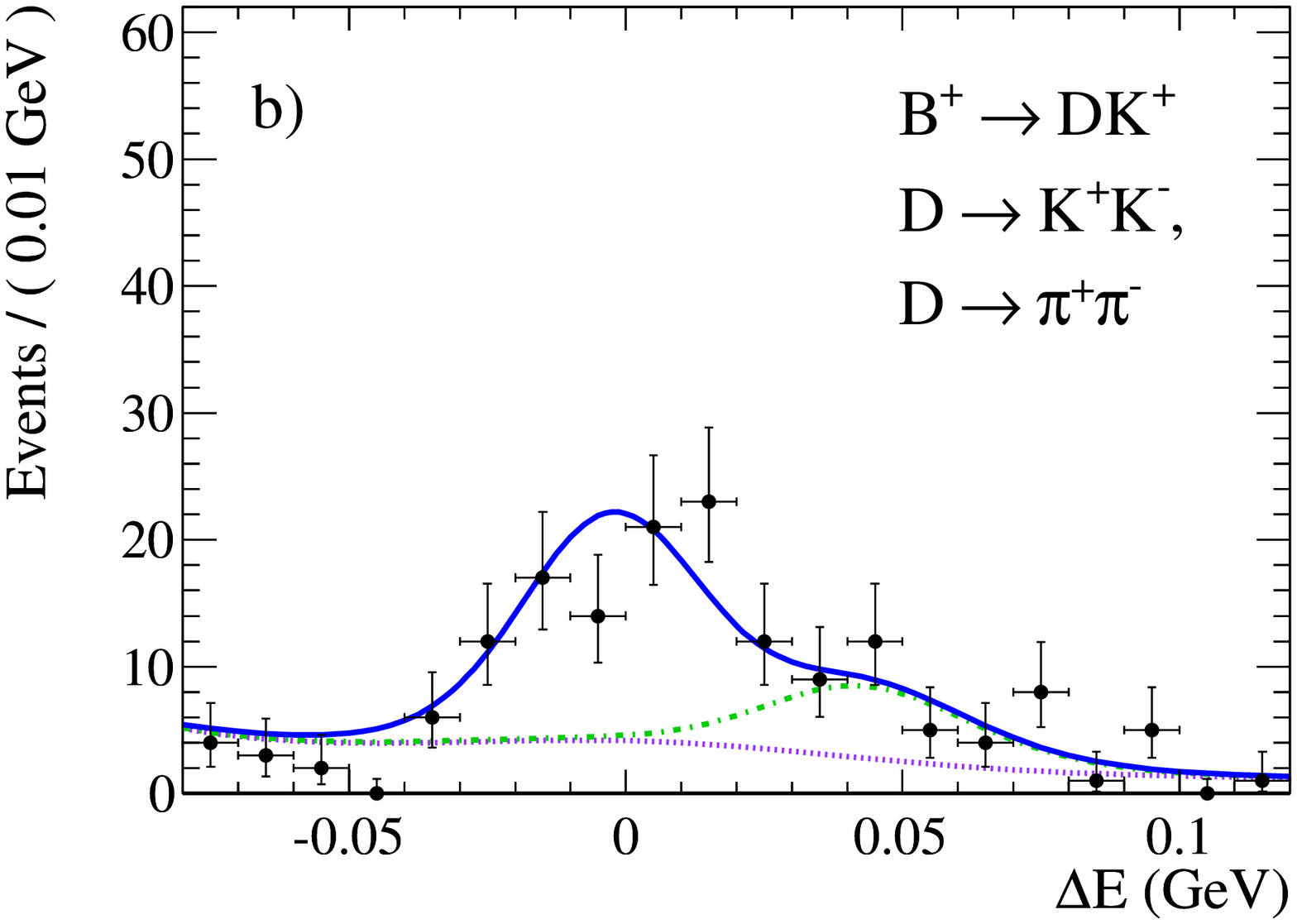}
\caption{\de projections of the fits to the data: (a) $B^-{\to}D_{\CP+}K^-$, 
(b) $B^+{\to}D_{\CP+}K^+$.
The curves are the full PDF (solid, blue), and 
$B{\to}D\pi$ (dash-dotted, green) stacked on the remaining backgrounds
(dotted, purple).
%The region between the solid and the dash-dotted
%lines represents the $B{\to}DK$ contribution.
%We show the subsets of the data sample in which the track 
%$h$ from the $B$ decay is identified as a kaon. 
We require candidates to lie inside a signal-enriched region:
$0.2<\fis<1.5$, $5.275<\mes<5.285\gevcc$, charged particle from the $B$ 
passing kaon identification criteria.
}
\label{fig:GLW_deltae_fits}
\end{figure}

The signal yields, from which the partial decay rates are determined,
are obtained from maximum likelihood fits to \mes, \de and \fis.
An example is shown in Fig.~\ref{fig:GLW_deltae_fits}.
We identify about 500 $B^\pm\to DK^\pm$ decays with \CP-even $D$ final states 
and a similar amount of $B^\pm\to DK^\pm$ decays with \CP-odd $D$ final states.
We measure $A_{\CP+} = \AcppVal\pm\AcppErrStat\pm\AcppErrSyst$ and 
and $A_{\CP-} = \AcpmVal\pm\AcpmErrStat\pm\AcpmErrSyst$, respectively, where
the first error is the statistical and the second is
the systematic uncertainty.
The parameter $A_{\CP+}$ is different from zero with a significance of 
3.6 standard deviations, constituting evidence for direct \CP
violation. We also measure $R_{\CP+} = \RcppVal\pm\RcppErrStat\pm\RcppErrSyst$ 
and $R_{\CP-} = \RcpmVal\pm\RcpmErrStat\pm\RcpmErrSyst$.

\begin{figure}[!htb]
\centering
\includegraphics[width=0.48\columnwidth,height=5cm]{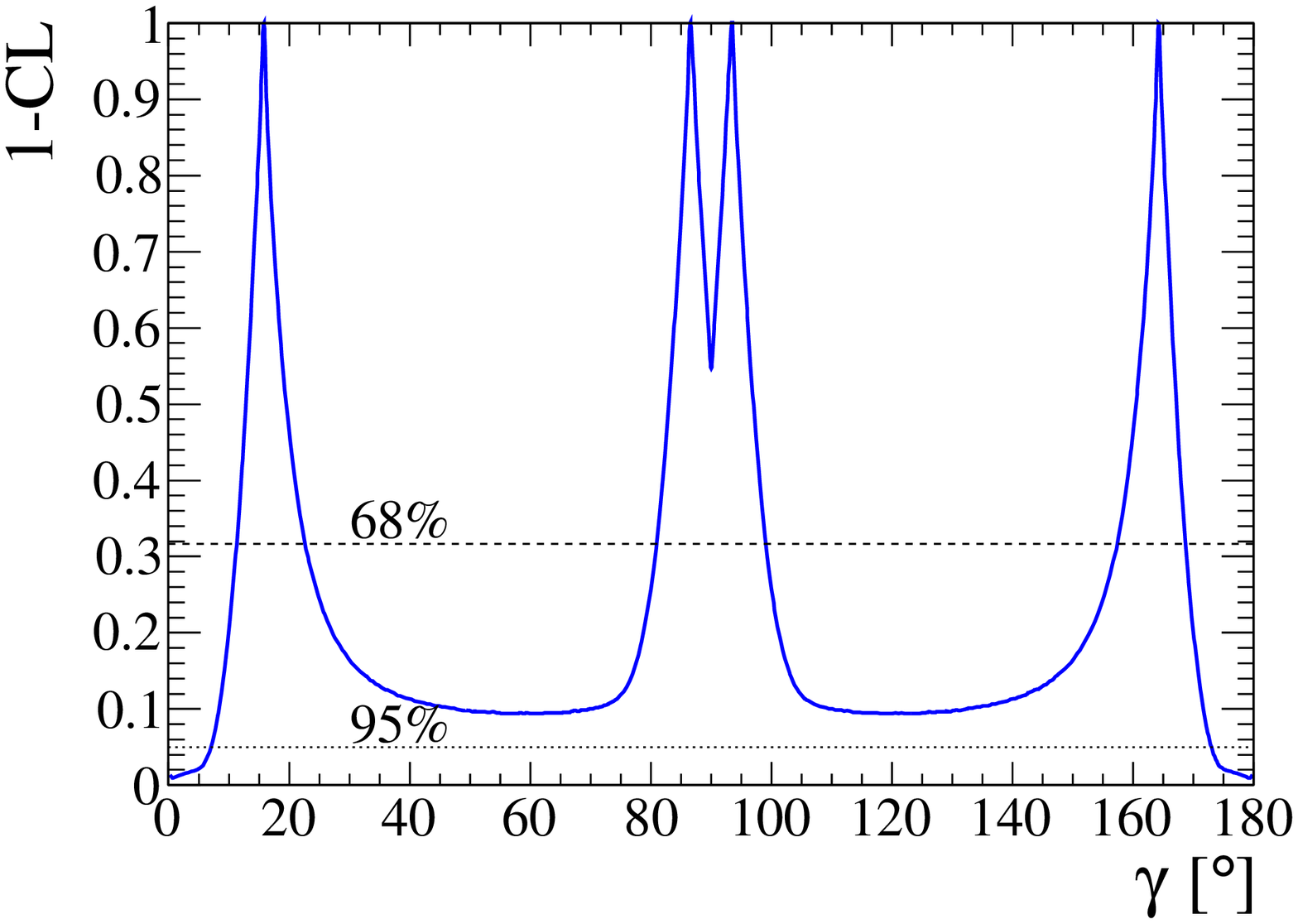}
\includegraphics[width=0.48\columnwidth,height=5cm]{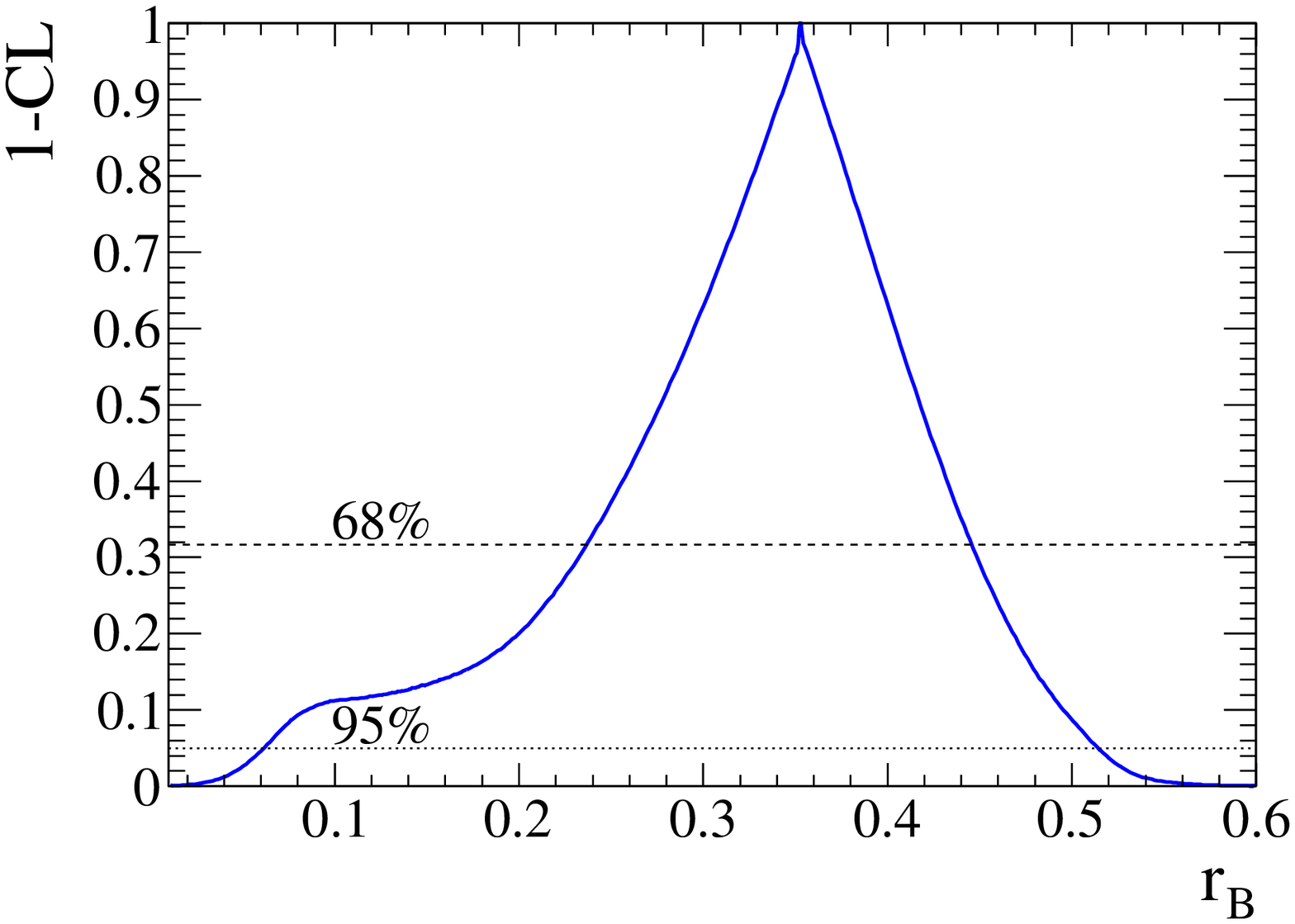}
\caption{1-CL as a function of $\gamma$ mod $180\degrees$ (left) and $r_B$ (right) from the $B{\to}DK$ GLW study.}
\label{fig:GLW_r_and_gamma}
\end{figure}

Using a frequentist technique, including statistical and systematic 
uncertainties, we obtain 
$\rLoA<r_B<\rHiA$ ($\rLoB<r_B<\rHiB$) and, modulo $180\degrees$,
$\dgLoAi\degrees<\gamma<\dgHiAi\degrees$ or 
$\dgLoAj\degrees<\gamma<\dgHiAj\degrees$ or
$\dgLoAk\degrees<\gamma<\dgHiAk\degrees$
($\dgLoBi\degrees<\gamma<\dgHiBi\degrees$) 
at the 68\% (95\%) confidence level (Fig.~\ref{fig:GLW_r_and_gamma}).
To facilitate the combination of these measurements with the results
of the Dalitz-plot analysis, we exclude the $D\to \KS\phi$, $\phi\to K^+K^-$
channel from this analysis -- thus removing events common to the two measurements --
and express our results in terms of the variables $x_{\pm}$ using 
$x_\pm = \frac{1}{4}\left[R_{\CP+}(1\mp A_{\CP+}){-}R_{\CP-}(1\mp A_{\CP-})\right]$. We find:
$x_+ = \xpValNoKsPhi\pm\xpErrStatNoKsPhi\pm\xpErrSystNoKsPhi$ and 
$x_- = \xmValNoKsPhi\pm\xmErrStatNoKsPhi\pm\xmErrSystNoKsPhi$,
in good agreement with the results from the Dalitz-plot analysis.

\section{ADS method: $B^\pm\to D^{(*)}K^\pm$, $D\to K^\pm\pi^\mp$}
We reconstruct $B^\pm{\to}DK^\pm$ and $D^{*}K^\pm$
($D^{^*}{\to}D\gamma$ and $D\pi^0$), followed by $D$ decays to both
the doubly-Cabibbo-suppressed $D^0$ final state $K^+\pi^-$ and the
Cabibbo-allowed final state $K^-\pi^+$, which is used as normalization
and control sample~\cite{bib:babar_ADS_dk}. 
Final states with opposite-sign kaons are produced from the
interference of the CKM favored $B$ decay followed by the doubly
Cabibbo-suppressed $D$ decay and the CKM- and color- suppressed $B$
decay followed by the Cabibbo-allowed $D$ decay, and the \CP
asymmetries may be potentially very large. On the other hand, their
overall branching fractions are very small ($O(10^{-7})$) and
background suppression is crucial.
The three branching fraction ratios ($R_{ADS}$) between $B$ decays
with opposite-sign and same-sign kaons and the three charge
asymmetries ($A_{ADS}$) in $B$ decays with opposite-sign kaons provide
six observables that can be used, together with the measurements by
$c$- and $B$-factories of the amplitude ratio $r_D$ and the strong
phase difference $\delta_D$ between the two $D$ decay amplitudes, to
determine $\gamma$ (with a 4-fold discrete ambiguity) and the two sets of $r_B,\delta_B$~\cite{bib:ADS}.

\begin{figure}[!htb]
\centering
\includegraphics[width=0.48\columnwidth]{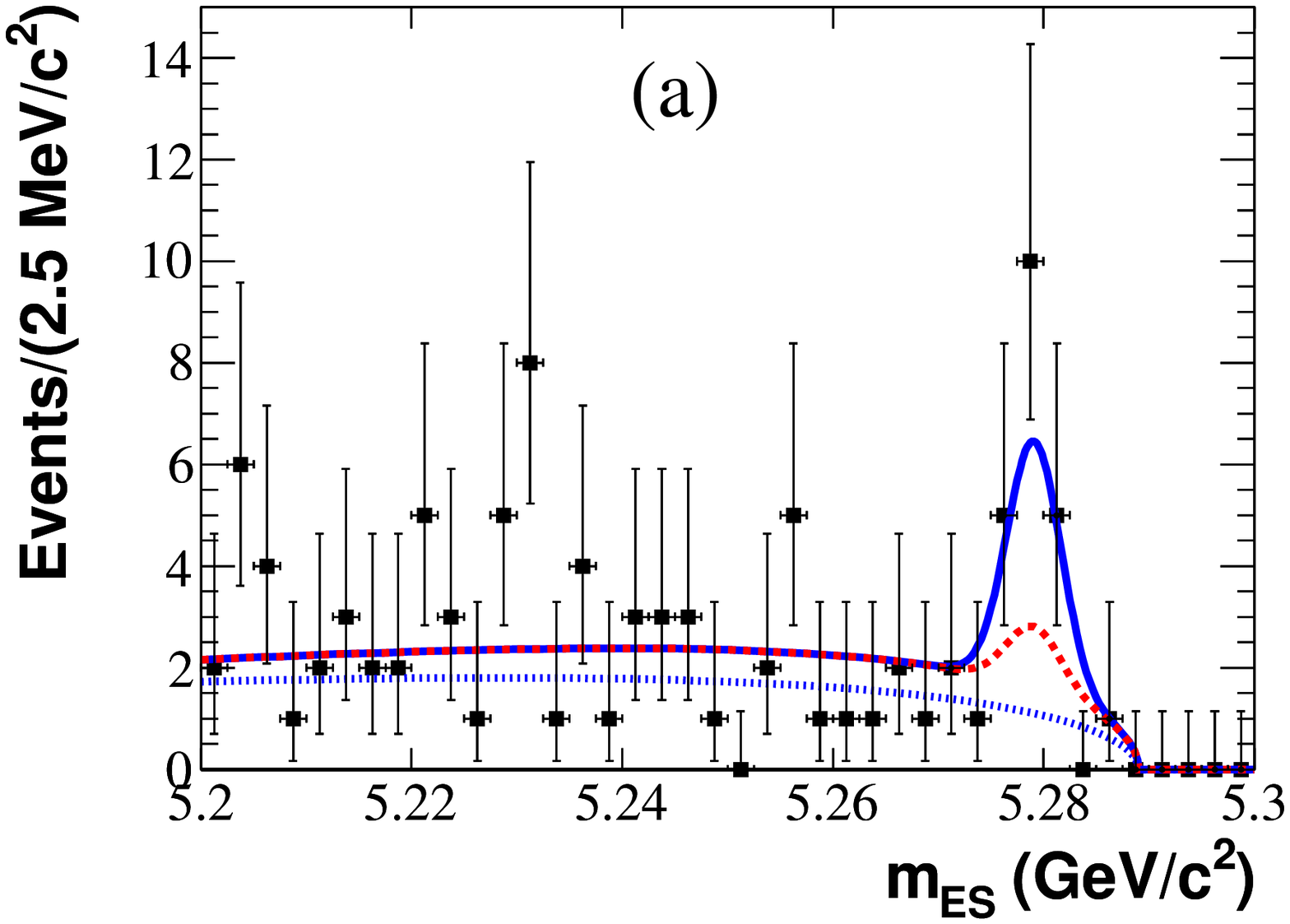}
\includegraphics[width=0.48\columnwidth]{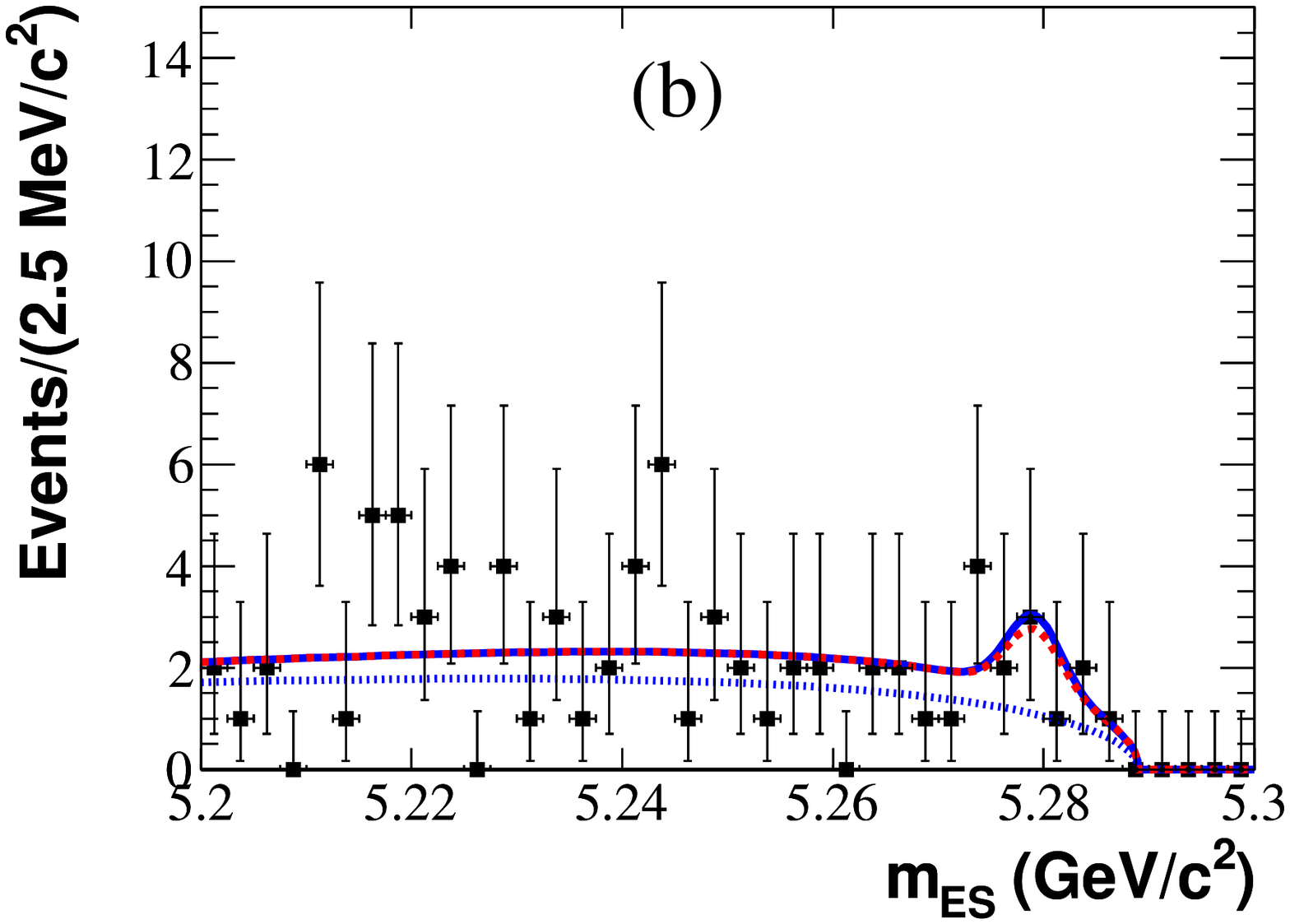}
\caption{
\mes projection of the fit to the data for the $B^\pm\to DK^\pm$,
$D\to K^\mp\pi^\pm$ decays, for samples enriched in
signal $(NN>0.94)$, for (a) $B^+$ and (b) $B^-$ 
candidates.
The curves represent the fit projections for
signal plus background (solid), the sum of all background
components (dashed), and the $q\bar q$ background only (dotted).}
\label{fig:dKpresultsads}
\end{figure}

\begin{figure}[!htb]
\centering
\includegraphics[width=0.48\columnwidth]{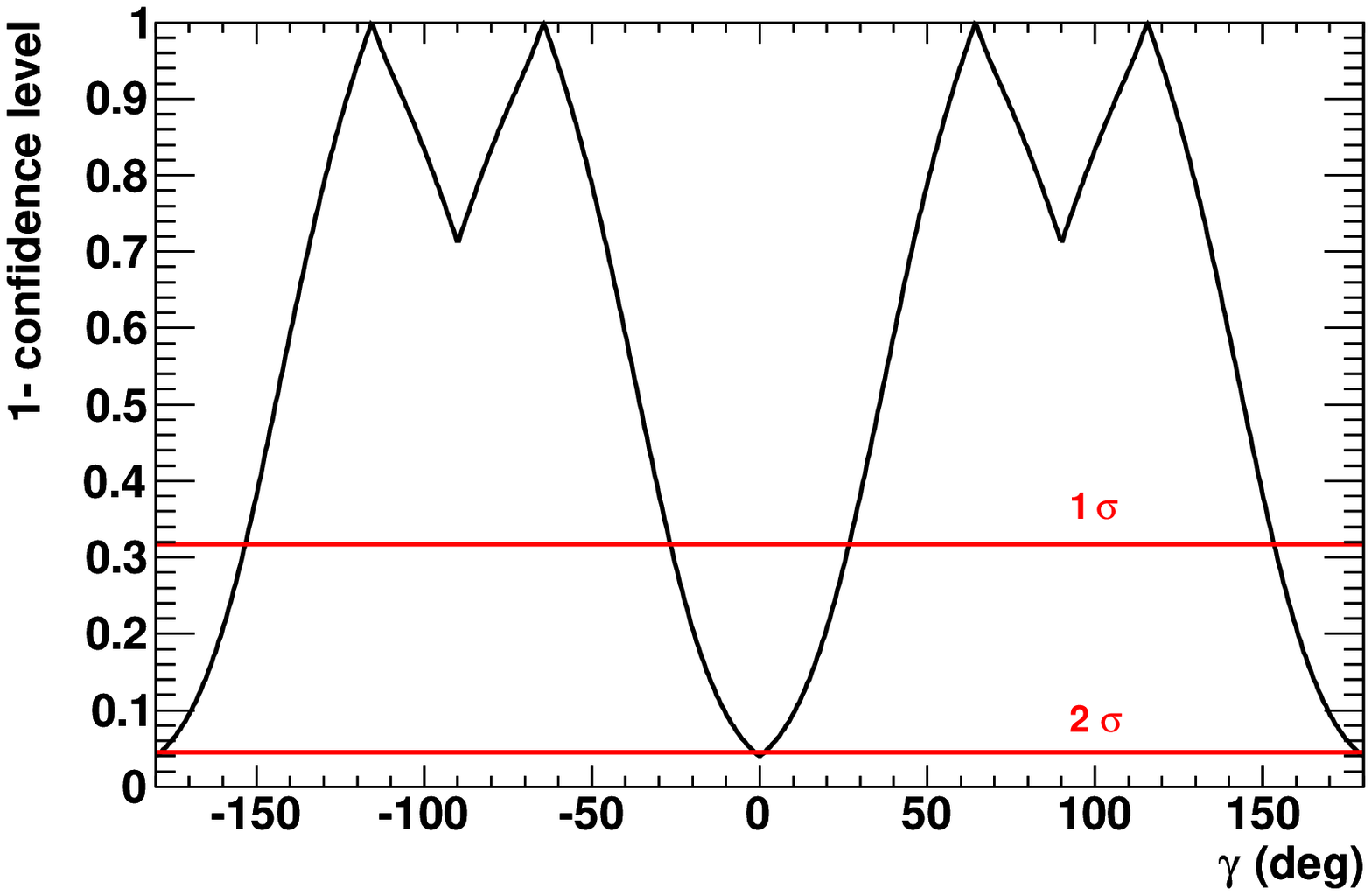}
\includegraphics[width=0.48\columnwidth]{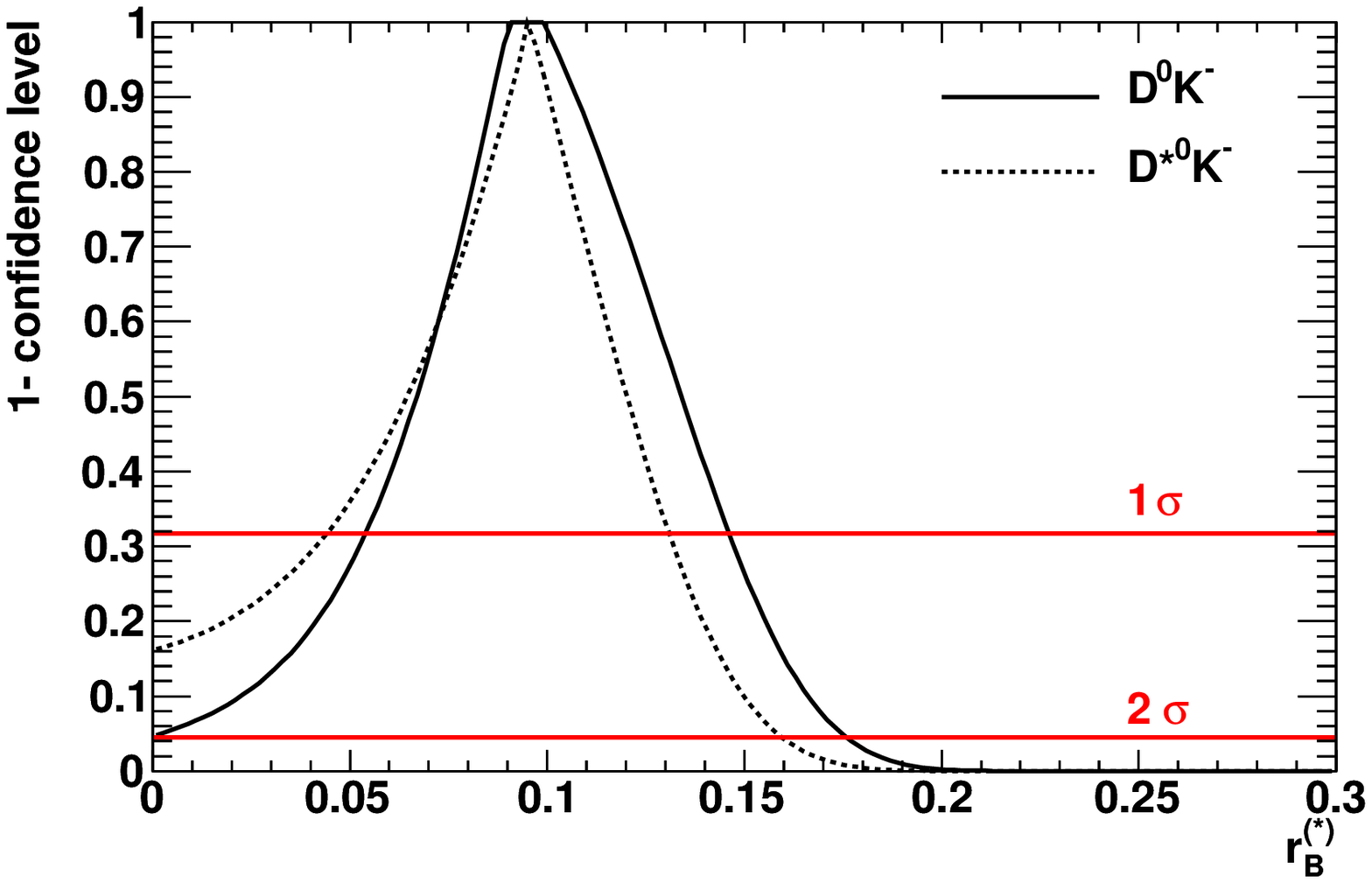}
\caption{1-CL as a function of $\gamma$ (left) and $r_B$ (right) from the $B{\to}D^{(*)}K$ ADS study.}
\label{fig:ADS_r_and_gamma}
\end{figure}

%The analysis method is first applied to $B^-\to D^{(*)}\pi^-$, where the
%$D$ decays into the Cabibbo favored ($K^-\pi^+$) and doubly suppressed modes ($K^+\pi^-$).
The yields are determined from fits to \mes and $NN$ (Fig.~\ref{fig:dKpresultsads}).
We see indications of signals for the $B\to DK$
and $B\to D^{*}_{D\pi^0}K$ opposite-sign modes, with significances of
$2.1\sigma$ and $2.2\sigma$, respectively. The measured branching fration ratios 
are $R_{ADS}^{DK} = (1.1\pm 0.5 \pm 0.2)\times 10^{-2}$
and $R_{ADS}^{D\pi^0K} = (1.8\pm 0.9 \pm 0.4)\times 10^{-2}$.
The \CP asymmetries are large, $A_{ADS}^{DK} = -0.86 \pm 0.47 \
^{+0,12}_{-0.16}$ and $A_{ADS}^{D\pi^0K} = +0.77 \pm
0.35\pm 0.12 $. We see no evidence of opposite-sign $B\to D^{*}_{D\gamma}K$ decays,
and measure $R_{ADS}^{D\gamma K} = (1.3\pm 1.4\pm
0.8 )\times 10^{-2}$ and $A_{ADS}^{D\gamma K} = +0.36 \pm 0.94\
^{+0.25}_{-0.41}$. From these results we infer
$r_B^{DK^\pm} = 0.095^{+0.051}_{-0.041}$,
$r_B^{D^*K^\pm} = 0.096^{+0.035}_{-0.051}$ and 
$54\degrees<\gamma<83\degrees$ (Fig.~\ref{fig:ADS_r_and_gamma}).

\section{Conclusion}
The full \babar\ dataset has been exploited to measure the CKM angle $\gamma$ 
in several $B^\pm\to D^{(*)}K^{(*)\pm}$ decays using three alternative techniques.
A coherent set of results on $\gamma$ and on the hadronic parameters 
characterizing the $B$ decay amplitudes has been obtained. 
The central value for $\gamma$, around $70\degrees$, is consistent with 
indirect determinations from the CKM fits. 
We attained a precision on $\gamma$ around $15\degrees$, and confirm the theoretical 
expectations of significant suppression ($r_B\approx 0.1$) of the $b\to u$ 
mediated decay amplitud with respect to the $b\to c$ one.
Finally, two direct CP violation evidences at the level of $3.5\sigma$ have been
observed.

\end{document}